\documentclass[10pt,conference]{IEEEtran}
\IEEEoverridecommandlockouts
\usepackage{cite}
\usepackage{amsmath,amssymb,amsfonts}
\usepackage{algorithmic}
\usepackage{graphicx}
\usepackage{textcomp}
\usepackage{xcolor}

\usepackage{ifpdf}
\usepackage{cite}
\usepackage{placeins}

\usepackage{balance}
\usepackage{algorithmic}
\usepackage{array}
\usepackage{mdwmath}
\usepackage{mdwtab}
\usepackage{amsmath}
\usepackage{algorithm}
\usepackage{xcolor}
\usepackage{eqparbox}
\usepackage{comment}
\usepackage{mathtools}
\usepackage{booktabs}
\usepackage{multirow}
\usepackage{xspace}
\usepackage{subcaption}

\usepackage{etoolbox,environ}

\IEEEoverridecommandlockouts\IEEEpubid{\makebox[\columnwidth]{978-1-6654-0601-7/22/\$31.00 ~\copyright~2022 IEEE \hfill} \hspace{\columnsep}\makebox[\columnwidth]{ }}

\newtoggle{shortver}
\toggletrue{shortver}
 
\NewEnviron{shortver_env}
  {\iftoggle{shortver}{\BODY}{}}

\def\BibTeX{{\rm B\kern-.05em{\sc i\kern-.025em b}\kern-.08em
    T\kern-.1667em\lower.7ex\hbox{E}\kern-.125emX}}
\begin{document}

\title{Dynamic-Deep: Tune ECG Task Performance and Optimize Compression in IoT Architectures}

\author{
\IEEEauthorblockN{Eli Brosh}
\IEEEauthorblockA{\textit{Computer Science Department} \\
\textit{Reichman University}\\
 Herzliya, Israel\\
eliahubrosh@gmail.com}
\and
\IEEEauthorblockN{Elad Wasserstein}
\IEEEauthorblockA{\textit{Computer Science Department} \\
\textit{Reichman University}\\
 Herzliya, Israel\\
elad.wasserstein@post.idc.ac.il}
\and
\IEEEauthorblockN{Anat Bremler-Barr}
\IEEEauthorblockA{\textit{Computer Science Department} \\
\textit{Reichman University}\\
 Herzliya, Israel\\
bremler@idc.ac.il}
}
\maketitle

\begin{abstract}
Monitoring medical data, e.g., Electrocardiogram (ECG) signals, is a common application of Internet of Things (IoT) devices. Compression methods are often applied on the massive amounts of sensor data generated prior to sending it to the Cloud to reduce the storage and delivery costs. A lossy compression provides high compression gain (CG), but may reduce the performance of an ECG application (downstream task) due to information loss. Previous works on ECG monitoring focus either on optimizing the signal reconstruction or the task's performance. Instead, we advocate a self-adapting lossy compression solution that allows configuring a desired performance level on the downstream tasks while maintaining an optimized CG that reduces Cloud costs.

We propose Dynamic-Deep, a task-aware compression geared for IoT-Cloud architectures. Our compressor is trained to optimize the CG while maintaining the performance requirement of the downstream tasks chosen out of a wide range. In deployment, the IoT edge device adapts the compression and sends an optimized representation for each data segment, accounting for the downstream task's desired performance without relying on feedback from the Cloud. 
We conduct an extensive evaluation of our approach on common ECG datasets using two popular ECG applications, which includes heart rate (HR) arrhythmia classification. We demonstrate that Dynamic-Deep can be configured to improve HR classification F1-score in a wide range of requirements. One of which is tuned to improve the F1-score by 3 and increases CG by up to 83\% compared to the previous state-of-the-art (autoencoder-based) compressor. Analyzing Dynamic-Deep on the Google Cloud Platform, we observe a 97\% reduction in cloud costs compared to a no compression solution.

To the best of our knowledge, Dynamic-Deep is the first end-to-end system architecture proposal to focus on balancing the need for high performance of cloud-based downstream tasks and the desire to achieve optimized compression in IoT ECG monitoring settings.

\end{abstract}

\section{Introduction}
Internet of Things (IoT) devices are widely used to monitor and send sensor data to the Cloud for centralized storage and execution of downstream tasks. For example, hospitals use IoT medical devices to constantly monitor Electrocardiogram (ECG) signals, the patient heart's activity over time. Several downstream tasks make use of ECG signals: For alerting the staff of HR arrhythmias~\cite{rajpurkar2017cardiologist} or extracting indicative features (e.g. R-R peaks \cite{vijayarangan2020rpnet}) for the purpose of heart disease diagnostics

\begin{figure}[t]
  \includegraphics[scale=.45]{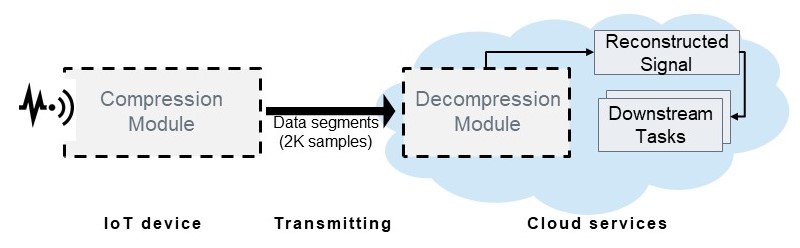}
  \caption {Typical architecture of a modern IoT-based medical monitoring system with data compression.}
  \label{fig:HL_IoT_architecture}
\end{figure}

Such medical monitoring settings generate large amounts of continuous sensor data to be sent to the Cloud. Transmitting the raw signal would imply power-hungry devices and high processing and storage Cloud costs. Therefore, an effective data compression scheme is required to reduce the transmission and storage requirements. An efficient compression module is typically deployed on the device to accommodate settings with low power resource-limited IoT devices, while a decompression module is deployed in the Cloud. Compression gain (CG) is usually used to evaluate such data compression schemes by calculating the division sizes of the original representation against the compressed representation. Fig. \ref{fig:HL_IoT_architecture} presents an end-to-end data flow in a Cloud-based monitoring system: compressed data segments are prepared for transmission, and the data is decompressed back to raw ECG signal in the Cloud for further processing and analysis.


\begin{shortver_env}
\begin{table}
\centering
\caption{Evaluate HR arrhythmia classification and reconstruction tasks on 602 test segments after compression}
\label{tab:comparison}

\resizebox{.51\textwidth}{!}
{%
\begin{tabular}{lllll}
\hline
\multicolumn{1}{|l|}{Compression Method}                      & \multicolumn{1}{l|}{CG}             & \multicolumn{1}{l|}{\begin{tabular}[c]{@{}l@{}}Avg. signal \\ reconstruction\\ error {[}\%{]}\end{tabular}} & \multicolumn{1}{l|}{\begin{tabular}[c]{@{}l@{}}HR classification\\ F1-score (Precision,Recall)\end{tabular}} & \multicolumn{1}{l|}{\begin{tabular}[c]{@{}l@{}}Violation \\ of upper\\ bound** {[}\%{]}\end{tabular}} \\ \hline
\multicolumn{1}{|l|}{LZ77 (Lossless)}       & \multicolumn{1}{l|}{2.7}            & \multicolumn{1}{l|}{0}                                                                               & \multicolumn{1}{l|}{\textbf{0.87*} (0.87,0.87)}                                                       & \multicolumn{1}{l|}{\textbf{0}}                                                                       \\ \hline
\multicolumn{1}{|l|}{CAE (SOTA)}            & \multicolumn{1}{l|}{32.25}          & \multicolumn{1}{l|}{\textbf{2.73}}                                                                   & \multicolumn{1}{l|}{0.20 (0.32,0.15)}                                                                 & \multicolumn{1}{l|}{10.21}                                                                            \\ \hline
\multicolumn{1}{|l|}{\textbf{Dynamic-Deep \#1}} & \multicolumn{1}{l|}{{48.31}} & \multicolumn{1}{l|}{3.81}                                                                            & \multicolumn{1}{l|}{0.73 (0.72,0.74)}                                                                 & \multicolumn{1}{l|}{\textbf{0}}                                                                       \\ \hline
\multicolumn{1}{|l|}{\textbf{Dynamic-Deep \#2}} & \multicolumn{1}{l|}{\textbf{52.2}} & \multicolumn{1}{l|}{4.2}                                                                            & \multicolumn{1}{l|}{0.69 (0.698,0.697)}                                                                 & \multicolumn{1}{l|}{\textbf{0}}                                                                       \\ \hline
\multicolumn{5}{l}{\begin{tabular}[c]{@{}l@{}}* The HR arrhythmia classification F1-score (Precision,Recall) results when applied to uncompressed\\      data segments\\ ** Percentage of data segments that experience classification error above  0.75\end{tabular}}                                       
\end{tabular}%
}
\end{table}
\end{shortver_env}

General ECG compression techniques belong to one of two: lossless or lossy. Lossless compression methods preserve the signal's complete information but tend to achieve low compression gain (CG). In contrast, lossy techniques (that are a natural fit to IoT settings) obtain high CG at the expense of losing information, and hence reduce the performance of downstream tasks. For example, we show in Table \ref{tab:comparison} that a common lossless compression,  LZ77~\cite{welch1984technique}, achieves a CG of 2.7 (with STD of 0.08) on widely used ECG  datasets with 602 test data segments of 2K samples each, rendering the scheme inappropriate in many scenarios. On the other hand, utilizing a state-of-the-art (SOTA) lossy compression scheme, based on a convolutional autoencoder (CAE)~\cite{yildirim2018efficient} tuned to a high fixed CG of 32, leads to significant performance degradation in signal reconstruction and heart rate classification when evaluating on the CinC test dataset~\cite{clifford2015physionet}.

In Cloud settings, it is desired to have a lossy compression scheme sensitive to Cloud costs that maintains a high performance for multiple downstream tasks executed on the decompressed data. Previous ECG works focus either on compressing schemes optimized for signal reconstruction (and can result in low downstream task performance) or models optimized for extracting or classifying ECG features (and not necessarily lend themselves to effective compression)~\cite{hong2020opportunities}. One approach to bridge the gap between the two is by controlling the tradeoff between compression level and tasks' performance.  

In this paper, we aim to provide a lossy compression method that allows the system administrator (admin) configuring a desired performance level, namely an upper bound on downstream tasks' error (that can be also tuned to be zero), while maintaining an optimized CG to reduce Cloud costs. Moreover, the method needs to be resource efficient to fit in an IoT device without adding significant overhead. 

We propose Dynamic-Deep (see Fig.~\ref{fig:propsed_Architecture}), a task-aware variable-rate compression. Our analysis reveals a low correlation between reconstruction error and the (tested) downstream task's error. The low correlation may be caused by the objective difference, continuously tracking the signal vs. capturing sporadic events. Hence, Dynamic-Deep jointly optimizes CG and the tasks' performance level. It learns to choose the optimal CG dynamically for each data segment sent to the Cloud out of multiple compression levels, each implemented by a single convolutional autoencoder. The dynamic behavior allows the scheme to achieve impressive CG. With such a design, one can directly specify the desired performance level of the tasks, rather than provide a general hyper-parameter to control a rate quality tradeoff, as often done in variable rate compression schemes.

A straightforward approach to extend an existing CAE to support dynamic operation is to choose the optimal level of compression based on the reconstructed signal. Such a solution requires feedback from the downstream tasks and thus is not a fit for an IoT setting where the compression and decompression modules are decoupled. To address this, we extend a classical CAE architecture also to learn predicting the downstream tasks' feedback as part of the compressed representation. This prediction eliminates the feedback loop between the IoT devices and the Cloud but still assures an optimal compressed representation that meets desired performance of downstream tasks. To further reduce the computational costs of our model, we apply standard (deep learning) architecture optimization techniques. We believe that our work may be the first to propose a dynamic compression scheme with a tunable downstream task error that easily matches the design requirements for IoT medical applications. 

\begin{shortver_env}
\begin{figure}[t]  \centerline{\includegraphics[scale=.41]{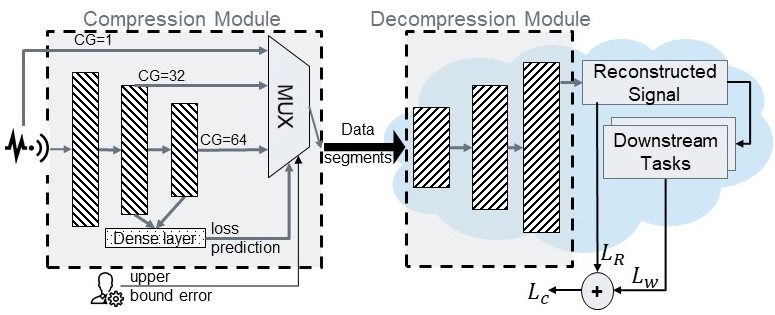}}
  \caption []{Schematic architecture of Dynamic-Deep. The compression and decompression modules use stacked convolutional layers with downsample and upsample layers, respectively, to achieve multiple compression levels\footnotemark. The encoder uses a dense layer to predict the task error for each level.
}
  \label{fig:propsed_Architecture}
\end{figure}
\footnotetext{Modern implementations of ECG applications are encoder-based. Our design uses a dedicated encoder for compression, decoupled from the downstream encoder, to allow a lightweight compressor implementation and support cases where the downstream task internals are not available to the admin due to business or IP restrictions.}
\end{shortver_env}

We conducted a comprehensive evaluation of our proposed method using two complex real-life ECG applications, R-R peaks detection (RPNet~\cite{vijayarangan2020rpnet}) and HR arrhythmia classification~\cite{rajpurkar2017cardiologist} with widely used ECG datasets. 
The dataset includes 50 hours ECG recordings with different types of labeled arrhythmia annotated by medical experts. The test set has 602 data segments with 8.9K samples each while 40\% contains abnormal events. Our compression method can be configured to a wide range of downstream task performances. One of which can increase HR arrhythmia classification F1-score by 3 times, and the average CG by up to 83\% (with a maximum CG of 64), compared to a SOTA CAE compression with a fixed CG~\cite{yildirim2018efficient} (Table \ref{tab:comparison} compare arbitrary working knobs against SOTA CAE and lossless method).  These improvements stem from training the compressor with the downstream task loss feedback (yielding better F1-score) and applying a dynamic compression approach (yielding better CG). Dynamic-Deep successfully preserved the desired error bound for all data segments; Whereas SOTA CAE violates the HR arrhythmia classification task's error bound for 10.21\% of the data segments. Here, the classification task's error is defined as the categorical cross entropy (CCE) loss, a common loss function for classification models. Further, the memory footprint of our solution is 67\% lower than that of the CAE. In addition, we deployed our solution on the Google Cloud Platform to study its real-world Cloud costs.
Decompressing data traffic equivalent to that of a small-mid-sized hospital (200 beds), we  
observed reductions in Cloud expenses 
by up to 97\% compared to that of non-compressed data traffic.

Our method can be easily extended to incorporate additional downstream tasks by adding the corresponding loss functions. Moreover, while we present the results of Dynamic-Deep on medical ECG data, our method is general, and thus can be extended to support other sensor types and potentially other domains by simply adapting the compressor modules~\footnote{Source code available at https://github.com/eladwass/Dynamic-Deep}.

The paper outlined the following: Section \ref{sec:rework} discusses  related works on compressing ECG signals.  
Sections \ref{sec:motivations}-\ref{sec:method} explain the motivation, proposed solution and implementation details. Section \ref{sec:experiemnts} shows evaluation experiments and conclusions in \ref{sec:conclusion}.

\section{Related Work}
\label{sec:rework}

Existing compression techniques were adapted into medical IoT environments, typically to fit the low power requirements ~\cite{ukil2015iot},~\cite{djelouat2018compressive}. Such methods are split into lossless~\cite{welch1984technique} and lossy~\cite{djelouat2018compressive} categories. Lossless achieves low CG on signals, such as ECG~\cite{koski1997lossless}, while lossy compression is highly efficient (in reducing storage requirements), and thus fit for IoT sensing.

{\bf Transform-based compression.} A common approach for lossy compression is transform-based, which seeks to preserve the crucial parts of the signal's representation in the transformed domain. Few notable examples are Fourier transform ~\cite{reddy1986ecg},  Wavelet transform ~\cite{chen1993ecg} or the Cosine transform ~\cite{ahmed1974discrete}, each with its own domain transformation preference. The transformed representation of ECG data is often sparse, and thus preserving the right parts of the representation enables one to get a reconstruction signal of potentially high fidelity~\cite{djelouat2018compressive},~\cite{cheng2007data}. In~\cite{ukil2015iot} a proposal for a dynamic scheme that adopts the threshold on the  representation size is suggested.  The main drawback of the transform-based approach is the use of a predefined domain transformation, which may not lead to the highest CG for the desired reconstruction level.

{\bf Neural network (NN) based compression.}  NNs automate the process of searching for an optimal domain transformation for the compressed representation. Auto-Encoders (AE), a family of NNs, were extensively studied~\cite{hinton2006reducing}, \cite{goroshin2013saturating} and shown to effectively learn an expressive-yet-efficient representation of ECG segments, and thus provide a higher CG than transform-based compression methods~\cite{ukil2015iot}. Recent work~\cite{yildirim2018efficient} uses a convolutional AE with 27 layers to achieve SOTA compression results. 
In general, AE architectures have a single fixed compression level, and thus are limited in the achievable CGs.

\emph{Variable-rate compression.} 
Recent works suggested NN architectures with multiple compression levels allow balancing between CG and reconstruction quality~\cite{choi2019variable}~\cite{akbari2020learned}. However, they require to define in advance a rate control parameter. Additionally, they focus their evaluation on reconstruction quality, thus there is no guarantee on downstream tasks performance. Whereas Dynamic-Deep is tuned to a desired downstream tasks' performance and optimizes CG for each data segment automatically rather than manual parameter changes.

{\bf Task-aware compression.}
Recent NN-based data compression architectures consider the joint performance of signal reconstruction and single downstream task, such as Person keypoint detection~\cite{pinkham2020algorithm} or 3D point cloud classification tasks~\cite{dovrat2019learning}. Such methods perform better with respect to downstream tasks performance and compression gain than its standalone single-task counterparts. However, none of them introduced performance analysis of multiple downstream tasks. Dynamic-Deep applies such an approach for the ECG domain and studies how to tune a multiple-level task-aware compressor to optimize overall CG results.

\section{Motivation For Balancing CG And Downstream Tasks Performance}
\label{sec:motivations}

Lossy compression methods, such as SOTA CAE~\cite{yildirim2018efficient}, are based on an architecture with a single fixed compression level. However, a fixed compression level may not necessarily satisfy a desired bound on the task's error for every data segment.

\begin{figure}[t]
    \centering
    \includegraphics[scale=.33]{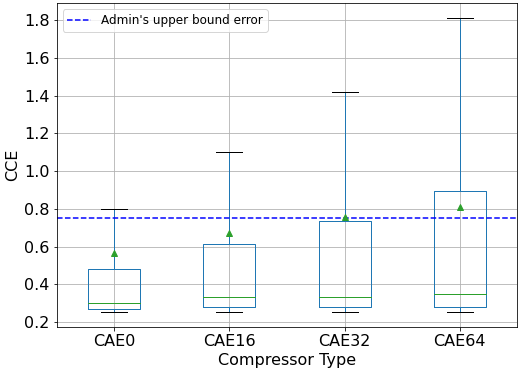}
    \caption[]{Categorical cross entropy loss quartiles for HR arrhythmia classification over test ECG data segments (of size 602) 
    }
    \label{fig:boxplotApp_loss}
\end{figure}

Fig. \ref{fig:boxplotApp_loss} presents the CCE quartiles across ECG tested segments for different compression levels: 64, 32, 16 and none. (See section~\ref{sec:Datasets} for details on the task and dataset used).  Denote CAExx as an extended SOTA CAE implementation with a CG of xx, and let CAE0 represent no compression. We note  that  non  zero  losses  may  occur  in uncompressed operations due to the inherent error of the HR detection model.

Let an example scenario be when the admin bound the HR arrhythmia classification loss (CCE) to 0.75, none of the fixed compressors can satisfy this bound for all segments, as shown in Fig.~\ref{fig:boxplotApp_loss}. To meet the bound of 0.75 we can apply an approach of 2-compression levels by combining a single compressor and no compression. For instance, with a single compressor CAE32, 75\% of data segments meet the upper bound, hence the rest 25\% remain non compressed. Such a setting yields an average CG of 24. Additionally, we can leverage higher CG as long as those meet the upper bound error of desired downstream task. For instance, CAE64 meets the upper bound for approximately more than 70\% of data segments. Therefore, increasing to 3-compression levels allows more compression possibilities with potential of higher CG. Back to the example meeting the bound of 0.75, approximately 75\% of data segments can be compressed with a high CG of 64 or 32 while the rest 25\% remain uncompressed, reaching higher CG of 48.31. 

Balancing CG and the downstream tasks' error is challenging, as the compressor needs the task's error feedback for every level of compression to optimize for the highest CG. But, in IoT settings, downstream tasks are decoupled from the compressor and located in the Cloud. We explain how to manage and provide the error feedback in the next section.

\section{Dynamic-Deep High Level Design}
\label{sec:proposed}
To receive feedback from the downstream tasks, we propose a solution that extends the classical CAE architecture with multiple compression levels in a task-aware fashion. 
Dynamic-Deep consists of the following (as shown in Fig.~\ref{fig:propsed_Architecture}):
\begin{enumerate}
\item \emph{Compression Module}: set of encoders that compress every raw data segment into multiple compression level representations. To reduce the need for actual feedback from the downstream tasks, we employ a \emph{dense layer} trained to predict the downstream tasks' weighted error for each compression level. We choose the highest feasible CG based on the pre-configured upper bound error and the feedback prediction.
   
 \item \emph{Decompression Module}: set of decoders that accept multiple representations of different compression levels and reconstruct the data segment.

\end{enumerate}
For optimizing the joint performance of downstream tasks and reconstructed tasks, the training phase requires differentiable downstream tasks (e.g. NN-based). Nonetheless, after the training phase, the reconstructed signal (from the pre-trained compressor) allows executing additional downstream tasks not necessarily differentiable.
 We found a low correlation between reconstruction error and downstream tasks' error (see Section~\ref{sec:training}). We therefore designed  Dynamic-Deep to predict the downstream tasks' error rather than the reconstruction error as a proxy for the error feedback.

\section{Dynamic-Deep Implementation Details}
\label{sec:method}

\subsection{Multiple Compression Levels}
The following architectural changes were made to extend the CAE32 to support 3-compression levels (64, 32 or no compression). 
We chose 3-compression levels based on experiments comparing the performance (and footprint) of different numbers of compression levels, including CAE16. We obtained the highest impact with 3-compression levels (Full implementation details in~\cite{dynamic-deeptechreport}). Denote conv(x,y) as a convolutional layer with x number of filters, kernel size of y, and Upsample(z) as an upsample layer with size z. To support CG of 64, convolutional layers conv(64,7) and conv(1,3) were added after CAE32 encoder's layer number 12,  yielding an output shape of (31,1). 

A decoder adapter transforms the compressed representation of 64 to input the CAE32's decoder. The adapter has four layers: conv(16,7), conv(32,3), Upsample(2), conv(32,3), Upsample(2) with an output size of (124,32). The output of the adapter is the input of layer 18 in CAE32.  Each compressed representation has its \emph{dense layer} supporting the downstream tasks' prediction. The dense layer's output shape is (1,1) with a ReLU activation.

\subsection{Minimizing Memory Footprint For IoT Support}
\label{sec:mini}
Deploying compressing modules on lightweight IoT devices requires a resource-constrained implementation. We use two encoders to support 3-compression levels, which results in a  1.06MB memory footprint for the extended SOTA CAE. We use several techniques to reduce the model's memory size:
\begin{enumerate}
    \item Deep learning compression techniques: 
    The number of learnable parameters of the CAE's 10th layer is reduced by decreasing the number of filters and kernel size. We compensate and preserve compression performance by preserving the receptive field using convolutional layers with stride=2 instead of pooling layers~\cite{cheng2017survey}. Note that the encoder's 10th layer is used in all compression levels.
    \item Sharing layers: Each encoder (CAE32, CAE64) share the same learnable parameters, therefore memory and computation are reused to avoid linear memory increase for each compression level~\cite{han2015deep}. Only the last encoder's layers of each compression level are computed uniquely. 

\end{enumerate}
Table \ref{tab:reduction_memory} summarizes the resulting memory size when
Applying the above techniques. It decreases the number of parameters to only 84K parameters which are 67\% fewer parameters to the straightforward 3-compression level CAE's encoder.

\begin{table}[b]
\centering
\caption{Compression Module's Memory minimization}
\label{tab:reduction_memory}
\resizebox{0.45\textwidth}{!}{%
\begin{tabular}{|l|l|l|}
\hline
                      & \begin{tabular}[c]{@{}l@{}}Number of\\ parameters\end{tabular} & \begin{tabular}[c]{@{}l@{}}memory\\ size {[}KB{]}\end{tabular} \\ \hline
Original 2-compression level (CAE based)              & 133K                                                           & 532   
\\ \hline
Original 3-compression level (CAE based)              & 266K                                                           & 1064   
\\ \hline
Deep learning compression techniques          & 168K                                                           & 678                                                            \\ \hline
Sharing layers (\textbf{Dynamic-Deep 3-compression level})  & \textbf{83K}                                                   & \textbf{332}                                                   \\ \hline
\end{tabular}%
}
\end{table}

\subsection{Combined Loss Function}
\label{sec:loss}
Dynamic-Deep has three loss functions accumulated to a combined loss function.
The \emph{reconstruction loss} $L_R$, calculates the distance between each sample  in the original data segment $X$ and the reconstructed data segment $\hat{X}$ over $M$ samples. 
\begin{equation}
L_{R} = \frac{1}{M}\sum_{i=0}^{M-1}\frac{|X[i]-\hat{X}[i]|}{X[i]}*100
\label{eq:loss}
\end{equation}
\normalsize
The \emph{downstream task weighted error} $L_w$, accumulates the downstream tasks' loss functions. Let $t_i$ denote task $i$, and $L_{t_i}$ and $w_i$ denote its loss function and the weighted (scaling) factor of the loss function, respectively.  
\begin{equation}
L_{w} =  \sum_{i} w_i*L_{t_i}
\label{eq:weightedLoss}
\end{equation}
\normalsize
The \emph{combined loss function} $L_{c}$, combines the reconstruction loss $L_R$ with the downstream tasks' weighted error $L_w$. $w_0$ scales the $L_R$ to balance between reconstruction performance and downstream tasks performances. 
\begin{equation}
L_{c} = w_0*L_{R}  +  L_{w}
\label{eq:combinedLosses}
\end{equation}
\normalsize
Finally, Dynamic-Deep learns to predict $L_w$ using the \emph{mean squared error (MSE)}.

\subsection{Training}
\label{sec:training}
We observe a low correlation between the tested downstream tasks and the reconstruction errors across a wide range of the weighting factors $w_i$ (see tech report~\cite{dynamic-deeptechreport}). Hence, training a downstream task in isolation on the reconstructed signal may result in limited performance.  We thus train Dynamic-Deep in three phases, where the second phase is repeated for each additional downstream task.  First, we train the compressor to optimize the reconstruction task (see eq.~\eqref{eq:loss}). Here, we use the  MIT-BIH dataset by applying the preprocessing described in~\cite{yildirim2018efficient}.
Second, we fine-tune with cascaded downstream tasks loss, including reconstruction loss (see eq.~\eqref{eq:combinedLosses}). The downstream tasks' models' weights are frozen and trained for additional 20 epochs with the CinC dataset by applying the preprocessing described in~\cite{rajpurkar2017cardiologist}. Finally, training the \emph{dense layer} to predict the downstream tasks' weighted error for additional 10 epochs. As before, the compressor is frozen to not penalize previous training phases.

All phases use the Keras API with TensorFlow 2.2 backend, an
Adam optimizer with a learning rate of 0.001, $\beta_1=0.9$, $\beta_2=0.999$, decay$=1e-5$, and a batch size of 32.

\begin{figure}[t]
\captionsetup[subfigure]{justification=centering}
    \centering
      \begin{subfigure}{0.37\textwidth}
          \includegraphics[width=\textwidth]{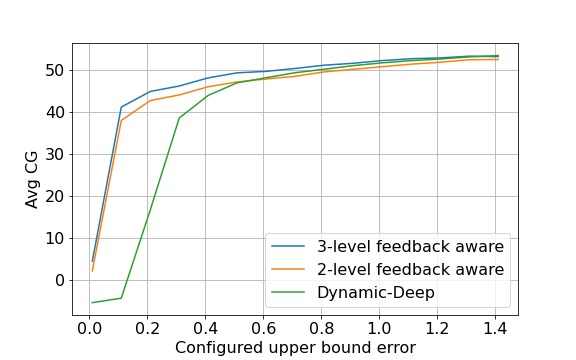}
  \caption {Average CG against pre-configurable bound}
  \label{fig:ANDREW_CG_proof}
          
      \end{subfigure}
      \begin{subfigure}{0.37\textwidth}
        \includegraphics[width=\textwidth]{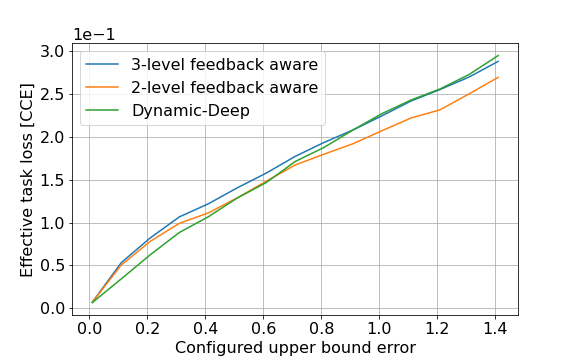}
         \caption {Effective loss against pre-configurable bound}
  \label{fig:ANDREW_EffectiveLOSS_proof}
      \end{subfigure}
      \caption{Benchmark on HR arrhythmia classification task }
\label{fig:andrew_dynamicDeep}
\end{figure}

\section{Experimental Results}
\label{sec:experiemnts}

\subsection{Datasets and Downstream Tasks}
\label{sec:Datasets}
\textbf{Datasets}: used for training and evaluation:
\begin{itemize}
    \item \emph{MIT-BIH} \cite{mark1988bih}: used to evaluate ECG compression as it includes different types of noise patterns and various shapes of arrhythmic QRS complexes \cite{hong2020opportunities}. The benchmark contains 48 half-hour ambulatory ECG recordings yielding a data set comprising 4.8K ECG data segments. 
     \item \emph{CinC} \cite{clifford2015physionet}: Captured from the AliveCor ECG monitor and contains about 7K records with 8960 samples each. These records are annotated by medical experts to the following classes: Atrial Fibrillation(AF), noise, other rhythms or normal.  The test set has 602 data records where 40\% are abnormal events. Training set has 5.4K records. 
\end{itemize}

\textbf{Downstream tasks}: 
(NN based) architectures were chosen:
\begin{itemize}
    \item \emph{HR arrhythmias classification}: implemented using convolutional NN (CNN) \cite{rajpurkar2017cardiologist} as a classification task. The ground truth (GT) of this task uses labels from the CinC dataset. Those labels are annotated by medical experts. 
    \item \emph{R-R peak extraction(RPnet)}: implemented using NN  \cite{vijayarangan2020rpnet} as a regression task locating the position in time of the peak. The GT of this task (i.e. R-R peaks on raw ECG) is generated by running the RPnet NN on the raw ECG signals offered by the CinC dataset. Then each compression level is evaluated relatively to the GT. 
\end{itemize}
Both are commonly used tasks in real-world ECG applications.

\subsection{Task Awareness Evaluation}

We focus our evaluation on the Dynamic-Deep downstream tasks' predictions performed by the \emph{dense layer} in the compression module. We compare our predictions against two theoretical models, in which the downstream task feedback is available for the compressor:
\begin{enumerate}
    \item \emph{2-level feedback-aware}: the method has 2-compression levels of 64 or 1 (act as a lower bound).
    \item \emph{3-level feedback-aware}: the method has 3-compression levels of 64, 32 or 1 (act as an upper bound).
\end{enumerate}
Each model executes the downstream tasks for every data segment and measures the error at each compression level. Then, they choose the highest compression that meets the configured upper bound error. If none meets the upper bound,  the no compression level is chosen. Note that such a method is not applicable in typical IoT settings since the feedback is not readily available at the edges. We evaluated Dynamic-Deep vs. the theoretical models above  considering the following setups:
\begin{enumerate}
    \item \emph{Single downstream task awareness}: of HR arrhythmia classification or R-R peak extraction. Fig. \ref{fig:andrew_dynamicDeep} 
    shows that Dynamic-Deep follows the trends of the theoretical method successfully. Increasing the upper bound error increases the CG and the effective task loss and vice versa. For every configured upper bound Dynamic-Deep results with a lower effective task loss than the configured upper bound. Lastly, there is an improvement in CG for both tasks when increasing the number of compression levels from 2 to 3 (more details in~\cite{dynamic-deeptechreport}).

    \item \emph{Multiple downstream tasks' awareness}: of both R-R peak extraction and HR arrhythmia classification. Supporting multiple downstream tasks introduces a tradeoff on downstream task optimization. Dynamic-Deep uses a downstream task weighted error (see eq. \ref{eq:weightedLoss}), which can be viewed as a single task awareness. This setup achieves similar results as single downstream task awareness (see tech report for demonstrations~\cite{dynamic-deeptechreport}).

\end{enumerate}

\begin{shortver_env}Higher CG than 64 improved the theoretical feedback-aware
up to an average CG of 100, however, Dynamic-Deep had low
performance utilization of those levels (Full experimental results in tech report~\cite{dynamic-deeptechreport}).\end{shortver_env}

\subsection{Cloud Cost Reduction Analysis Using Dynamic-Deep}
Optimizing CG in IoT settings has a direct impact on reducing storage costs and networking bandwidth. For Cloud costs evaluation, we consider storage and computation (to account data decompression) costs since Cloud inbound traffic is usually free. We compare the following operational models:
\begin{enumerate}
    \item \emph{Dynamic-Deep}: IoT device sends a compressed representation to the Cloud side, where it is stored, or decompressed to allow downstream tasks' execution. 
    \item \emph{Dynamic-Deep with uncompressed}: IoT device sends both compressed and uncompressed representations. The compressed representation is stored while downstream tasks operate on the uncompressed representation. 
    
\end{enumerate}
We assume that a domain expert reviews some portion of $x\%$ of historical sensor data, and accounts for the corresponding overhead, the cost of fetching data from storage and decompressing it, in both models. 

\begin{figure}[t]
  \centering
  \includegraphics[scale=.36]{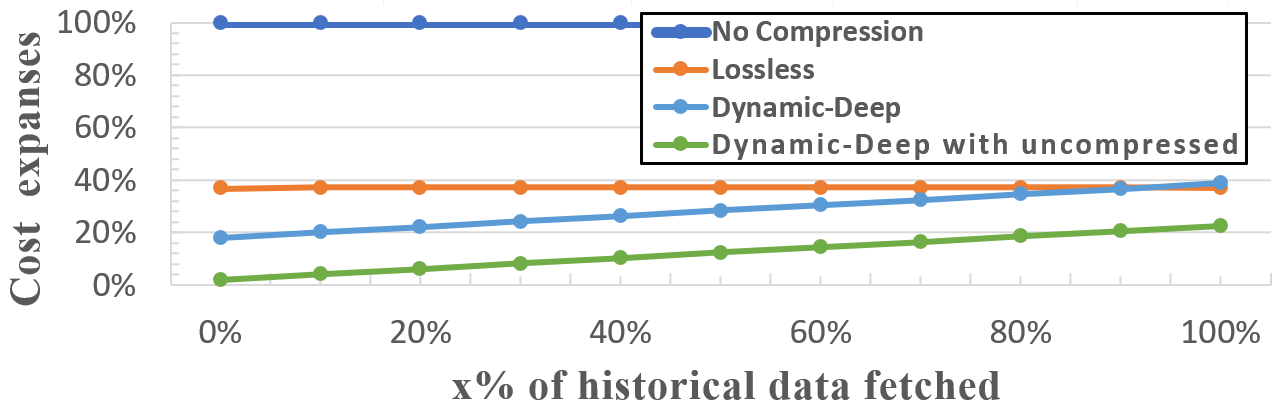}
  \caption {Yearly cost expenses compression comparison}
  \label{fig:costs}
\end{figure}

We ran these two models on Google Cloud Platform
~\cite{google} using ECG data segment traffic equivalent to a small-mid hospital (200 beds). We configured the upper bound error of HR arrhythmia classification to be 0.75 and received an average CG of 48.31.
We measured the computation expenses of our setup on an N1-Custom instance with 1 CPU, 2GB RAM, Intel Xeon 2.2GHz. Fig.~\ref{fig:costs} presents the measured results. Lossless compression reduces expenses by 63\% regardless of the specific architecture due to its low computation usage. \emph{Dynamic-Deep with uncompressed} architecture saves up to 97\% cost expenses compared to no compression solution and is more efficient than lossless even with 100\% data fetching.

\section{Conclusion}
\label{sec:conclusion}

We presented a self-adapted lossy compression architecture suitable for IoT networks that allow tuning ECG downstream tasks' performances and optimize the process of compressing ECG data segments using a variable-rate compression. The IoT device learns to predict the downstream task error to allow independent functionality from Cloud services. We successfully showed CG improvements on two types of downstream tasks against a SOTA CAE. Additionally we showed the method allows the practitioner to balance between desired performance and compression gain thus also controlling Cloud costs. Future work will extend the implementations to other domains.

\textbf{Acknowledgment} - We thank Guy Vinograd from bio-T for his comments, and deeply grateful to Elad Levy for his valuable feedback on model design and analysis.

\bibliographystyle{IEEEtranS}
\bibliography{sections/references}

\begin{thebibliography}{10}
\providecommand{\url}[1]{#1}
\csname url@samestyle\endcsname
\providecommand{\newblock}{\relax}
\providecommand{\bibinfo}[2]{#2}
\providecommand{\BIBentrySTDinterwordspacing}{\spaceskip=0pt\relax}
\providecommand{\BIBentryALTinterwordstretchfactor}{4}
\providecommand{\BIBentryALTinterwordspacing}{\spaceskip=\fontdimen2\font plus
\BIBentryALTinterwordstretchfactor\fontdimen3\font minus
  \fontdimen4\font\relax}
\providecommand{\BIBforeignlanguage}[2]{{%
\expandafter\ifx\csname l@#1\endcsname\relax
\typeout{** WARNING: IEEEtranS.bst: No hyphenation pattern has been}%
\typeout{** loaded for the language `#1'. Using the pattern for}%
\typeout{** the default language instead.}%
\else
\language=\csname l@#1\endcsname
\fi
#2}}
\providecommand{\BIBdecl}{\relax}
\BIBdecl

\bibitem{ahmed1974discrete}
N.~Ahmed, T.~Natarajan, and K.~R. Rao, ``Discrete cosine transform,''
  \emph{IEEE transactions on Computers}, vol. 100, no.~1, pp. 90--93, 1974.

\bibitem{akbari2020learned}
M.~Akbari, J.~Liang, J.~Han, and C.~Tu, ``Learned variable-rate image
  compression with residual divisive normalization,'' in \emph{2020 IEEE
  International Conference on Multimedia and Expo (ICME)}.\hskip 1em plus 0.5em
  minus 0.4em\relax IEEE, 2020, pp. 1--6.

\bibitem{chen1993ecg}
J.~Chen, S.~Itoh, and T.~Hashimoto, ``Ecg data compression by using wavelet
  transform,'' \emph{IEICE TRANSACTIONS on Information and Systems}, vol.~76,
  no.~12, pp. 1454--1461, 1993.

\bibitem{cheng2007data}
A.~F. Cheng, S.~E. Hawkins~III, L.~Nguyen, C.~A. Monaco, and G.~G. Seagrave,
  ``Data compression using chebyshev transform,'' 2007.

\bibitem{cheng2017survey}
Y.~Cheng, D.~Wang, P.~Zhou, and T.~Zhang, ``A survey of model compression and
  acceleration for deep neural networks,'' \emph{arXiv preprint
  arXiv:1710.09282}, 2017.

\bibitem{choi2019variable}
Y.~Choi, M.~El-Khamy, and J.~Lee, ``Variable rate deep image compression with a
  conditional autoencoder,'' in \emph{Proceedings of the IEEE/CVF International
  Conference on Computer Vision}, 2019, pp. 3146--3154.

\bibitem{clifford2015physionet}
G.~D. Clifford, I.~Silva, B.~Moody, Q.~Li, D.~Kella, A.~Shahin, T.~Kooistra,
  D.~Perry, and R.~G. Mark, ``The physionet/computing in cardiology challenge
  2015: reducing false arrhythmia alarms in the icu,'' in \emph{2015 Computing
  in Cardiology Conference (CinC)}.\hskip 1em plus 0.5em minus 0.4em\relax
  IEEE, 2015, pp. 273--276.

\bibitem{djelouat2018compressive}
H.~Djelouat, A.~Amira, and F.~Bensaali, ``Compressive sensing-based iot
  applications: A review,'' \emph{Journal of Sensor and Actuator Networks},
  vol.~7, no.~4, p.~45, 2018.

\bibitem{dovrat2019learning}
O.~Dovrat, I.~Lang, and S.~Avidan, ``Learning to sample,'' in \emph{Proceedings
  of the IEEE Conference on Computer Vision and Pattern Recognition}, 2019, pp.
  2760--2769.

\bibitem{dynamic-deeptechreport}
\BIBentryALTinterwordspacing
A.~B.-B. Elad~Wasserstein, Eli~Brosh, ``Dynamic-deep tech-report,'' 2022.
  [Online]. Available: \url{https://cutt.ly/bvekoKD}
\BIBentrySTDinterwordspacing

\bibitem{google}
\BIBentryALTinterwordspacing
Google. Machine types,compute engine documentation, google cloud. [Online].
  Available: \url{{cloud.google.com/compute/docs/machine-types}}
\BIBentrySTDinterwordspacing

\bibitem{goroshin2013saturating}
R.~Goroshin and Y.~LeCun, ``Saturating auto-encoders,'' \emph{arXiv preprint
  arXiv:1301.3577}, 2013.

\bibitem{han2015deep}
S.~Han, H.~Mao, and W.~J. Dally, ``Deep compression: Compressing deep neural
  networks with pruning, trained quantization and huffman coding,'' \emph{arXiv
  preprint arXiv:1510.00149}, 2015.

\bibitem{hinton2006reducing}
G.~E. Hinton and R.~R. Salakhutdinov, ``Reducing the dimensionality of data
  with neural networks,'' \emph{science}, vol. 313, no. 5786, pp. 504--507,
  2006.

\bibitem{hong2020opportunities}
S.~Hong, Y.~Zhou, J.~Shang, C.~Xiao, and J.~Sun, ``Opportunities and challenges
  of deep learning methods for electrocardiogram data: A systematic review,''
  \emph{Computers in Biology and Medicine}, p. 103801, 2020.

\bibitem{koski1997lossless}
A.~Koski, ``Lossless ecg encoding,'' \emph{Computer Methods and Programs in
  Biomedicine}, vol.~52, no.~1, pp. 23--33, 1997.

\bibitem{mark1988bih}
R.~Mark and G.~Moody, ``Mit-bih arrhythmia database directory,''
  \emph{Cambridge: Massachusetts Institute of Technology}, 1988.

\bibitem{pinkham2020algorithm}
R.~Pinkham, T.~Schmidt, and A.~Berkovich, ``Algorithm-aware neural network
  based image compression for high-speed imaging,'' in \emph{2020 IEEE
  International Conference on Artificial Intelligence and Virtual Reality
  (AIVR)}.\hskip 1em plus 0.5em minus 0.4em\relax IEEE, 2020, pp. 196--199.

\bibitem{rajpurkar2017cardiologist}
P.~Rajpurkar, A.~Y. Hannun, M.~Haghpanahi, C.~Bourn, and A.~Y. Ng,
  ``Cardiologist-level arrhythmia detection with convolutional neural
  networks,'' \emph{arXiv preprint arXiv:1707.01836}, 2017.

\bibitem{reddy1986ecg}
B.~S. Reddy and I.~Murthy, ``Ecg data compression using fourier descriptors,''
  \emph{IEEE Transactions on Biomedical Engineering}, no.~4, pp. 428--434,
  1986.

\bibitem{ukil2015iot}
A.~Ukil, S.~Bandyopadhyay, and A.~Pal, ``Iot data compression: Sensor-agnostic
  approach,'' in \emph{2015 Data Compression Conference}.\hskip 1em plus 0.5em
  minus 0.4em\relax IEEE, 2015, pp. 303--312.

\bibitem{vijayarangan2020rpnet}
S.~Vijayarangan, R.~Vignesh, B.~Murugesan, S.~Preejith, J.~Joseph, and
  M.~Sivaprakasam, ``Rpnet: A deep learning approach for robust r peak
  detection in noisy ecg,'' in \emph{2020 42nd Annual International Conference
  of the IEEE Engineering in Medicine \& Biology Society (EMBC)}.\hskip 1em
  plus 0.5em minus 0.4em\relax IEEE, 2020, pp. 345--348.

\bibitem{welch1984technique}
T.~A. Welch, ``A technique for high-performance data compression,''
  \emph{Computer}, no.~6, pp. 8--19, 1984.

\bibitem{yildirim2018efficient}
O.~Yildirim, R.~San~Tan, and U.~R. Acharya, ``An efficient compression of ecg
  signals using deep convolutional autoencoders,'' \emph{Cognitive Systems
  Research}, vol.~52, pp. 198--211, 2018.

\end{thebibliography}

\end{document}